\documentclass[aps,prd,twocolumn]{revtex4-2}

\usepackage[colorlinks]{hyperref}
\usepackage{amssymb,amsmath}
\usepackage{graphicx}

\begin{document}
\title{Migdal-type effect in the dark matter absorption process}
\author{V. A. Dzuba}\email{v.dzuba@unsw.edu.au}
\author{V. V. Flambaum}\email{v.flambaum@unsw.edu.au}
\author{I. B. Samsonov}\email{igor.samsonov@unsw.edu.au}
\address{School of Physics, University of New South Wales,
Sydney 2052, Australia}

\begin{abstract}
We propose a new mechanism of absorption of dark matter particles in atoms which resembles the Migdal effect of inelastic dark matter scattering. In this process, atom may be ionized upon absorption of a scalar particle through the scalar-nucleon Yukawa-type interaction. The crucial difference from the inelastic dark matter scattering on atoms is that the total energy of the particle, including its rest mass $mc^2$-term, is transferred to the electron. As a result, the emitted electron kinetic energy is about six orders in magnitude bigger than that in the dark matter scattering process. This absorption process allows one to probe dark matter particles with a relatively small mass, in the range from 1 to 100 keV, that cannot be detected in the scattering process. It is also possible to detect hypothetical scalar particles emitted from the Sun. We calculate absorption cross sections of this process in Na, Si, Ar, Ge, I, Xe, and Tl target atoms and extract limits on the scalar-nucleon interaction constant from null results of XENONnT experiment. 
\end{abstract}

\maketitle

\section{Introduction} 
Recent advancements in dark matter (DM) detection technology have witnessed a remarkable surge in sensitivity, propelling the field into an era of unprecedented precision. Notably, liquid noble gas detectors, exemplified by argon and xenon time-projection chambers, and solid state DM particle detectors based on NaI, Ge and Si crystals, have emerged as pivotal instruments in the pursuit of understanding the elusive nature of DM, see, e.g., \cite{ReviewDetectors} for a review. These detectors, originally conceived for the detection of Weakly Interacting Massive Particle (WIMP) DM candidates through nuclear recoil, have achieved sensitivities nearing those required for neutrino detection at certain energies. Intriguingly, their versatility extends beyond the traditional WIMP paradigm, as these detectors are also sensitive to direct electron recoil arising from DM-atom scattering events and atomic absorption of pseudoscalar, scalar and vector DM candidate particles (see e.g. \cite{el1,el2,el3,el4,el5,Bloch2020,Roberts2023}).

In this work, we investigate the viability of DM detectors in capturing the signatures of Yukawa-type scalar-nucleon interactions in atomic ionization by the scalar particle absorption. This process has similar kinematics to the scalar field absorption in atoms through the scalar-electron interaction (see, e.g., \cite{Bao2021}). On the other hand, it resembles the Migdal effect in the scattering process when atom is ionised due to a sudden nucleus motion caused by collision with a heavy DM particle \cite{Ibe2018}. The Migdal effect potentially allows one to probe the sub-GeV DM particles in liquid noble gas DM detectors.

Theoretically, the scalar particles are very well motivated DM candidates arising in different scenarios such as sgoldstino, dilaton, relaxion, moduli and Higgs-portal DM (see, e.g., \cite{SnowmassScalar} for a review). For light scalar fields, with mass $m_\phi\ll1$\,eV, a variety of detection techniques have been employed and proposed: atomic clocks \cite{AtClocks1,AtClocks2,AtClocks3,AtClocks4,AtClocks5,AtCloks6}, accelerometers \cite{Accelerometers}, resonant-mass detectors \cite{ResMass}, gravitational wave detectors \cite{GrWave}, laser and maser interferometry \cite{AtClocks2,Interferometry1,Interferometry2,Interferometry3}, atomic and molecular spectroscopy \cite{Antypas1,Oswald1,Tretiak1}, cavity resonators \cite{Cav}, atomic transitions in trapped atoms and molecules \cite{ArvanitakiMolecules,HFtransitions} and permanent magnets \cite{Magnet}. For heavier scalar fields, most of these approaches are, however, inefficient, and different techniques are needed. In this paper, we demonstrate that liquid noble gas detectors are suitable for probing the scalar-nucleon coupling constant if the scalar field mass falls in the range from 1 to 50 keV. 

Although we focus mainly on non-relativistic massive scalar DM particles, light ultrarelativistic scalar particles potentially produced in the Sun with $O$(keV) energy may also be absorbed by atoms through the scalar-nucleon interaction. We derive the general expression for cross section of absorption of scalar particles on atoms which holds both for massive and massless scalar field. In contrast with the corresponding cross section of direct electron ionization from Ref.~\cite{Bao2021}, the ionization due to the scalar-nucleon interaction is described by  dipole atomic matrix element while the contributions from other multipoles are negligible. We calculate this matrix element and the corresponding ionization cross section numerically for a variety of atoms of experimental interest: Na, Si, Ar, Ge, I, Xe, and Tl. In Appendix, the results of these calculations are tabulated, and an analytic fitting function for the absorption cross section is presented. More detailed numerical files are provided in Supplemental Material.

It is important to note that actual DM absorption rates in liquid noble gas and solid state detectors may significantly differ from the ones estimated with atomic matrix element calculations because of interatomic interactions and collective electron effects in solids. To address this problem, we derive a universal relation between the scalar field absorption cross section and the photoionization one. Making use of this relation, scalar DM particle absorption rate may be estimated in actual experiments once the photoionization cross section is measured for the detector's active medium. This relation is similar to the one for the axio-electric effect \cite{PospelovRitzVoloshin,AxioElectric}.


\section{The model}
Let us consider a free DM particle with mass $m_\phi$ described by a real scalar field $\phi$. Given one quanta of this particle in a unit volume, this field is represented by the following plane wave solution
\begin{equation}
    \phi = \frac1{\sqrt{2\omega}} [e^{i(\vec k\vec r - \omega t)} + e^{-i(\vec k\vec r - \omega t)}]\,,
    \label{phi}
\end{equation}
with dispersion relation $\omega = \sqrt{m_\phi^2 + \vec k^2}$ in natural units with $\hbar=c=1$. This field may have Yukawa-type interaction with Dirac spinor fields of nucleon $\psi_n$ and electron $\psi_e$
\begin{equation}
    {\cal L}_\text{int} = g_{\phi n} \phi \bar\psi_n \psi_n + g_{\phi e} \phi \bar\psi_e \psi_e\,,
    \label{Lint}
\end{equation}
with some small coupling constants $g_{\phi n}$ and $g_{\phi e}$. Here  $\bar\psi_{n,e} = \psi_{n,e}^\dag \gamma^0$ are Dirac conjugate spinors and $\gamma^0 = \left(
 \begin{smallmatrix}
     {\bf 1}_{2\times2} & 0 \\ 0 & -{\bf 1}_{2\times 2}
 \end{smallmatrix}
 \right)$. The Lagrangian (\ref{Lint}) corresponds to the following Hamiltonian 
\begin{equation}
    H_\text{int} = - (Q e^{-i\omega t} + Q^\dag e^{i\omega t})\,,
\end{equation}
where the operator
\begin{equation}
    Q = \frac{1}{\sqrt{2\omega}} \left( g_{\phi n}\gamma^0_{(n)}e^{i \vec k\vec r_n} + g_{\phi e}\gamma^0_{(e)}e^{i \vec k\vec r_e} 
    \right)
    \label{K}
\end{equation}
consists of two terms which depend on the electron $\vec r_e$ and the nucleon $\vec r_n$ variables, respectively. 

More generally, in Eq.~(\ref{Lint}) one can consider independent coupling constants to proton, $g_{\phi pr}$, and neutron, $g_{\phi ne}$. Upon averaging over a nucleus with mass number $A$ and charge number $Z$, these couplings contribute to $g_{\phi n}$ as $g_{\phi n} = [Zg_{\phi pr} + (A-Z)g_{\phi ne}]/A$.

In an atom, all nucleons interact with the background scalar field $\phi$ coherently, while the electrons in different shells may respond differently. Therefore, upon considering scattering of the scalar field on an atom, one has to average the operator (\ref{K}) over the nucleons and sum over the atomic electrons. As a result, the atomic transitions due to the interaction with the background scalar field $\phi$ are driven by the operator
\begin{subequations}
\begin{align}
    \bar Q &= Q_N + Q_e\,,\\
    Q_N &= \frac{g_{\phi n} A}{\sqrt{2\omega}} e^{i\vec k \vec r_N} \,,\\
    Q_e &= \frac{g_{\phi e}}{\sqrt{2\omega}} \sum_{i=1}^Z \gamma^0_{(i)}e^{i\vec k \vec r_i} \,,
\end{align}
\end{subequations}
where $\vec r_N$ is the position vector of the nucleus and $\vec r_i$ are positions of atomic electrons. Below, we will use this operator for calculation of the atomic ionization rate due to absorption of the scalar DM particles.


\section{Atomic ionization upon absorption of a scalar particle}
We consider a process of absorption of one quanta of the scalar field (\ref{phi}) with speed $v = k/\omega$ and ejection of one of the atomic electrons with kinetic energy in the interval $[{\cal E}_\text{f},{\cal E}_\text{f}+dE]$. 
The cross section of this process has the following general form
\begin{equation}
    d\sigma_\phi = 2\pi \frac{\omega}{k} \sum_{\text{I,F}} | \langle \text{F} |\bar Q | \text{I} \rangle|^2\delta(E_\text{F}-E_\text{I}-\omega)dE\,,
    \label{dsigma}
\end{equation}
where $|\text{I}\rangle$ and $|\text{F}\rangle$ are the initial and final atomic states with energies $E_\text{I} = \frac{p_\text{i}^2}{2m_\text{at}} + {\cal E}_\text{i}$ and $E_\text{F} = \frac{p_\text{f}^2}{2m_\text{at}} + {\cal E}_\text{f}$, respectively. Here $p_\text{i}$ and $p_\text{f}$ are the initial and final momenta of the atom, respectively, and ${\cal E}_\text{i}$ and ${\cal E}_\text{f}$ are the corresponding energies of the electronic cloud. In Eq.~(\ref{dsigma}), the sum runs over all final electronic configurations with energy ${\cal E}_\text{f}$, and averaging over all initial electronic configurations with energy ${\cal E}_\text{i}$ is assumed.

Similar to photoionization, the atomic recoil energy may be neglected, and the DM particle energy is predominantly spent for the atomic transition, $\omega \approx {\cal E}_\text{f} - {\cal E}_\text{i}$. Imposing the conservation of energy we find the total cross of this process
\begin{equation}
    \sigma_\phi = 2\pi \frac{\omega}{k}\sum_{\text{I,F}} | \langle \text{F} |\bar Q |\text{I} \rangle|^2\,.
    \label{w}
\end{equation}

The atomic states in the coordinate representation may be written as
\begin{subequations}
\label{states}
\begin{align}
    \langle \vec r_{e\,i},\vec r_N |\text{I}\rangle &= e^{-i\vec p_\text{i} \vec r_\text{at}} \psi_\text{i}(\vec r_{e\,i} - \vec r_N)\,,\\
    \langle \vec r_{e\,i},\vec r_N|\text{F}\rangle &= e^{-i\vec p_\text{f} \vec r_\text{at}} \psi_\text{f}(\vec r_{e\,i} - \vec r_N)\,,
\end{align}
\end{subequations}
where
\begin{align}
    \vec r_\text{at} &=\frac{m_N \vec r_N + m_e \sum_{i=1}^Z \vec r_{e\,i}}{m_\text{at}}\,,\\
    m_\text{at} &= m_N + Z m_e \approx A m_p\,.
\end{align}
Here $\vec r_{e\, i}$, $i=1,\ldots,Z$, denote position vectors of atomic electrons and $m_p$ is the nucleon mass. In Eq.~(\ref{states}), $\psi_\text{i}(\vec r_{e\,i} )\equiv \langle \vec r_{e\,i} |\text{i}\rangle$ and $\psi_\text{f}(\vec r_{e\,i})\equiv \langle \vec r_{e\,i} |\text{f}\rangle$ represent wave functions of the electron cloud while the factors $e^{-i\vec p_\text{i} \vec r_\text{at}}$ and $e^{-i\vec p_\text{f} \vec r_\text{at}}$ take into account the motion of the center of mass of the atom.

With the atomic wave functions (\ref{states}), the matrix elements in Eq.~(\ref{w}) may be explicitly represented as
\begin{subequations}
\label{MNe}
\begin{align}
    &M_\text{fi}^{(N)}\equiv \langle\text{F}| Q_N | \text{I} \rangle = \frac{g_{\phi n} A}{\sqrt{2\omega}} \int d^3r_N d\tau \, e^{i\vec q \vec r_\text{at} + i\vec k \vec r_N} \nonumber \\
    &\times \psi_\text{f}^*(\vec r_{e\,i}- \vec r_N)
    \psi_\text{i}(\vec r_{e\,i}-\vec r_N)\,,
    \\
    &M_\text{fi}^{(e)}\equiv \langle \text{F}| Q_e | \text{I} \rangle = \frac{g_{\phi e}}{\sqrt{2\omega}} \int d^3r_N d\tau \, e^{i\vec q \vec r_\text{at}} \nonumber \\
    &\times \psi_\text{f}^*(\vec r_{e\,i} - \vec r_N)
    \sum_{l=1}^Z e^{i \vec k \vec r_{e l}}\gamma^0_{(l)}
    \psi_\text{i}(\vec r_{e\,i}-\vec r_N)\,,
\end{align}
\end{subequations}
with $\vec q = \vec p_\text{f} - \vec p_\text{i}$ the momentum transfer and $d\tau = d^3 r_{e1}\ldots d^3r_{eZ}$ is the integration measure over all electronic coordinates. After shifting the electronic variables in the integrand, $\vec r_{e\,i} - \vec r_N \to \vec r_{e\,i}$, we perform the integration over $d^3r_N$ that yields the momentum conservation, $\vec q + \vec k =0$. As a result, the matrix elements (\ref{MNe}) are expressed in terms of the electronic wave functions,
\begin{subequations}
\begin{align}
    M_\text{fi}^{(N)} &= \frac{g_{\phi n} A}{\sqrt{2\omega}} \langle \text{f} | \exp\left( -i\frac{m_e}{m_\text{at}}\vec k \vec R \right) |\text{i}\rangle\,,   \label{Mphi1}\\
    M_\text{fi}^{(e)} &= \frac{g_{\phi e}}{\sqrt{2\omega}} 
    \langle \text{f} | \sum_{l=1}^Z e^{i\vec k (\vec r_{e\,l}-\vec R m_e/m_\text{at})}\gamma^0_{(l)}|\text{i}\rangle
    \nonumber\\
    &\approx \frac{g_{\phi e}}{\sqrt{2\omega}} 
    \langle \text{f} | \sum_{l=1}^Z e^{i\vec k \vec r_{e\,l}}\gamma^0_{(l)}|\text{i}\rangle\,,
    \label{Mphi2}
\end{align}
\label{Mphi}
\end{subequations}
where $\vec R = \sum_{i=1}^Z \vec r_{e\,i}$ is the sum of position vectors of all atomic electrons. In the last line of (\ref{Mphi2}) we neglected a small term $\vec R m_e/m_\text{at}$ as compared with $\vec r_{e\,l}$.

\begin{figure*}
    \centering
    \begin{tabular}{cc}
    \includegraphics[width=5cm]{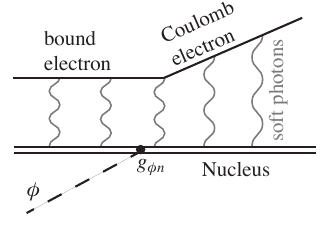} &
    \includegraphics[width=5cm]{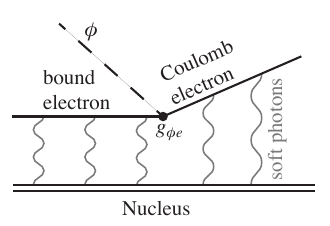}\\
    a & b
    \end{tabular}
    \caption{Diagrams representing atomic ionization by absorption of a scalar particle $\phi$ through (a) scalar-nucleon interaction and (b) scalar-electron vertex.}
    \label{fig:diagrams}
\end{figure*}

In the case when the final atomic state $|\text{f}\rangle$ corresponds to an ionized Coulomb electron, the transition amplitudes (\ref{Mphi}) may be visualized by diagrams in Fig.~\ref{fig:diagrams}.

In general, the coupling constants $g_{\phi n}$ and $g_{\phi e}$ are independent. However, if their values are comparable, $g_{\phi n}\sim g_{\phi e}$, the absorption through the nuclear recoil with the matrix element (\ref{Mphi1}) is strongly suppressed by the factor $m_e/m_\text{at}$, as compared with the direct electron ionization described by $M_\text{fi}^{(e)}$. The latter process was studied in Ref.~\cite{Bao2021} where limits on $g_{\phi e}$ were found from the results of the Xenon1T experiment \cite{XenonExcess}. In the present paper, we focus on the contributions from the matrix element (\ref{Mphi2}) which may be significant if $g_{\phi e}\ll g_{\phi n}$.

The matrix element (\ref{Mphi1}) contains a strong suppression factor $\frac{m_e}{m_\text{at}}\ll1$ under the exponent. As a result, for DM particles this matrix element is dominated by the dipole term,
\begin{equation}
    M^{(N)}_\text{fi}\approx -i \frac{g_{\phi n} A}{\sqrt{2\omega}} \frac{m_e}{m_\text{at}} 
    \langle \text{f} | \vec k \vec R |\text{i}\rangle\,.
\end{equation}
Substituting this matrix element into Eq.~(\ref{w}) and making use of the identity $|\langle \text{f}|\vec k \vec R |\text{i}\rangle |^2 = \frac13 k^2 |\langle \text{f}| \vec R |\text{i}\rangle |^2$ which holds for non-polarized atoms, we find the ionization cross section in the dipole approximation
\begin{equation}
    \sigma_\phi = \pi g_{\phi n}^2 \frac{m_e^2}{m^2_p} \frac{k}{3} \sum_{\text{i,f}} |\langle \text{f} | \vec R |\text{i}\rangle |^2\,.
    \label{sigma-phi}
\end{equation}

We stress that Eq.~(\ref{sigma-phi}) originates from the dipole approximation for the matrix element (\ref{Mphi1}). This approximation is applicable for $\langle \vec k \vec R\rangle m_e/m_\text{at}\ll1$. At large incoming particle momenta, the absorption is strongly dominated by inner $s$ and $p$ atomic shells. Therefore, $\langle R \rangle$ may be estimated as the radius of $1s$ shell, $\langle R \rangle\sim a_B/Z$, where $a_B$ is the Bohr radius. As a result, the cross section (\ref{sigma-phi}) is applicable up to a relatively high DM particle momenta 
\begin{equation}
    k\ll a_B^{-1}Z m_\text{at}/m_e\approx AZ\times 6.8\,\text{MeV}.
\end{equation} 

The cross section (\ref{sigma-phi}) represents the central result in the present work which allows us to study the absorption rate of the scalar particles in atoms and compare it with observational data. We calculate this cross section numerically for a variety of atomic targets of experimental interest: Na, Si, Ar, Ge, I, Xe, and Tl. The results of numerical calculations within the relativistic Hartree-Fock method are presented in tables in the Appendix. To facilitate further applications of these results, we also provide analytic fitting function for the absorption cross section. For Xe and Ar atomic targets, the numerically calculated cross sections are represented by orange dashed curves in Fig.~\ref{fig:CrossSections} which are in good agreement with the results obtained from experimental values of photoionization cross section considered below.


\section{Relation with photoionization cross section}
The scalar field absorption cross section (\ref{sigma-phi}) is expressed in terms of the standard atomic  dipole matrix element. The same matrix element gives the leading contribution to the photoionization cross section, see, e.g., \cite{Sobelman,PhotoCrossSection},
\begin{equation}
    \sigma_\gamma = \frac{4\pi^2\alpha}{3}\omega\sum_{\text{i,f}}
     | \langle \text{f} | \vec R | \text{i} \rangle |^2\,.
     \label{sigma-photo}
\end{equation}
As a result, we find the relation between the scalar field absorption cross section (\ref{sigma-phi}) and the photoionization one (\ref{sigma-photo}),
\begin{equation}
    \sigma_\phi = g_{\phi n}^2 \frac{m_e^2}{m^2_p}\frac{k}{\omega}\frac{\sigma_\gamma}{4\pi\alpha}\,.
\label{result}
\end{equation}
Note that the number of excited electrons produced by absorption of one particle may be bigger than one.

\begin{figure*}
    \centering
    \includegraphics[width=8cm]{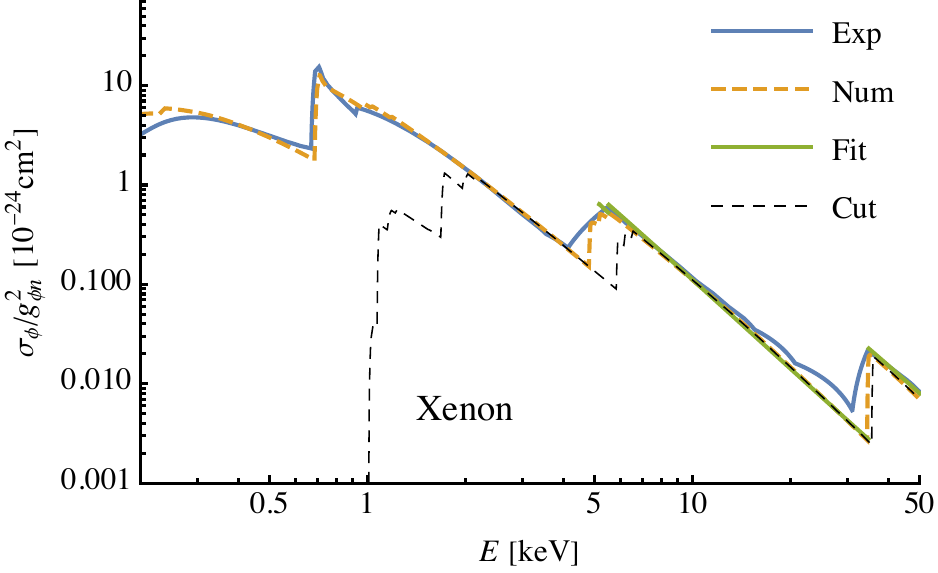}
    \includegraphics[width=8cm]{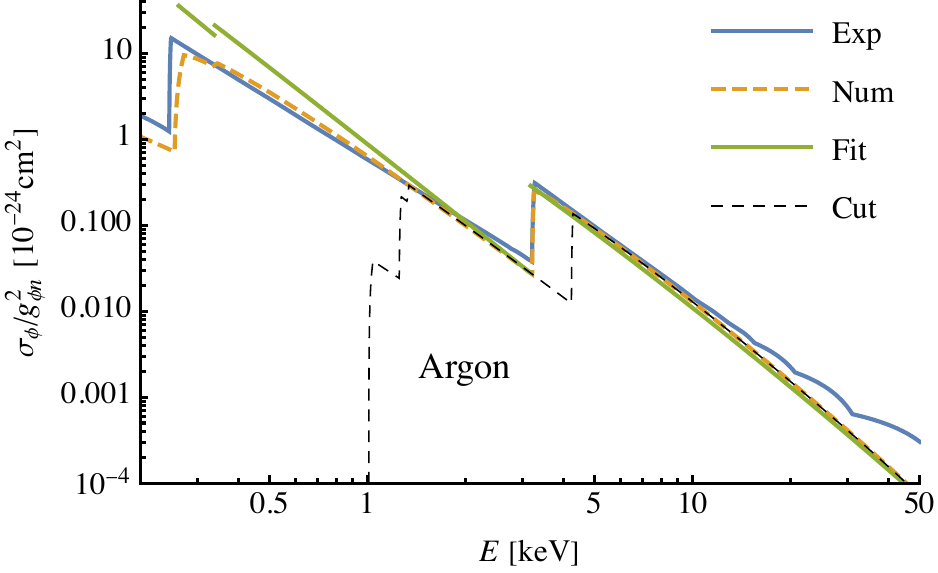}
    \caption{Absorption cross sections of a massless scalar field through scalar-nucleon interaction in Xe (left) and Ar (right) atomic targets. Blue curve corresponds to Eq.~(\ref{result}) with experimental values of the photoionization cross section from Refs.~\cite{Marr76,WEST1978}, orange dashed curve is obtained by direct numerical calculation of the matrix elements in Eq.~(\ref{sigma-phi}), black dashed curve is obtained by similar numerical calculations but assuming a detection threshold on ionized electron kinetic energy below 1 keV, and green curve is obtained with the use of the analytical fitting formula for ionization cross section (\ref{A11}). In the case of massive dark matter, the values of cross sections are to be reduced by the factor $v/c\sim 10^{-3}$. Note that at large energies the relation (\ref{result}) between the scalar and experimental photoionization  cross sections is violated since the photoinozation acquires high multipoles which do not contribute to the scalar ionization.}
    \label{fig:CrossSections}
\end{figure*}

The expression (\ref{sigma-photo}) for the photoionization cross section in the dipole approximation is applicable for photon wavelength greater than the typical size of the $1s$ atomic orbital, 
\begin{equation}
k\ll Z/a_B\approx Z \times 3.7\,\text{keV}.
\label{klimit}
\end{equation}
As a result, this bound sets the region of applicability of the relation (\ref{result}). In particular, for Xe atom, $Z=54$, Eq.~(\ref{result}) holds for scalar particle momenta $k\ll200$\,keV. This is well in the range of DM searching experiments with nuclear recoil in liquid Xe \cite{XENONnT,LZ,Panda}.

The photoionization cross section is calculated and measured experimentally to a high accuracy for a variety of atoms and molecules. In particular, for noble gases the value of this cross section is tabulated in Refs.~\cite{Marr76,WEST1978}. Using these data, in Fig.~\ref{fig:CrossSections} we present the plots of the absorption cross section (\ref{result}) of a massless scalar field in Ar and Xe targets. Note that in the case of massive scalar field the cross sections are similar, but are reduced by a factor $k/\omega=v/c$ which is of order $10^{-3}$ for galactic dark matter particles.

Although the relation (\ref{result}) is derived for atomic targets, it may be applied for liquid and solid state detectors as well. Making use of this relation, scalar DM particle absorption rate may be estimated in actual experiments once the photoionization cross section is measured for the detector’s active medium. Thus, Eq.~(\ref{result}) is a powerful tool for studying the scalar DM particle absorption with various detectors.


\section{Constraints from XENON-nT}
Consider massive DM particles virialized in the halo of our Galaxy with a speed distribution $f(v)$.  The rate of absorption of such particles on atoms with emission of electrons is
\begin{align}
    R &= \frac1{m_\text{at}}\frac{\rho_\text{DM}}{m_\phi}  \int \sigma_\phi(v) f(v) v dv \nonumber\\
    &=\frac{g_{\phi n}^2}{4\pi\alpha} \frac1{m_\text{at}}
\frac{\rho_\text{DM}}{m_\phi}\frac{m_e^2}{m_p^2}
\int \sigma_\gamma f(v)v^2 dv\,,
\label{rate1}
\end{align}
where in the second line we made use of the relation (\ref{result}). Note that the photoionization cross section $\sigma_\gamma$ depends on the DM particle velocity through the energy of the scalar field $E = m_\phi(c^2 + v^2/2)$. This dependence is very weak for non-relativistic DM particles. Hence, we obtain the following expression for the absorption rate of the scalar field:
\begin{equation}
    R = \frac{g_{\phi n}^2}{4\pi\alpha} \frac1{m_\text{at}}
\frac{\rho_\text{DM}}{m_\phi}\frac{m_e^2}{m_p^2}
\langle v^2 \rangle \sigma_\gamma(m_\phi)\,,
\label{rate2}
\end{equation}
where $\langle v^2\rangle = \int  v^2f(v)dv\approx 1.4\times 10^{-6}c^2$. Here we employed the DM particles velocity distribution function corresponding to the standard dark matter halo model from Refs.~\cite{formfactor,Foster2018}:
\begin{equation}
    f(v) = \frac{v}{\sqrt{\pi}v_0v_\text{obs}}e^{-(v+v_\text{obs})^2/v_0^2}(e^{4vv_\text{obs}/v_0^2}-1)\,.
\end{equation}
Here $v_0\approx 220$\,km/s is the speed in the local rotation curve and $v_\text{obs}\approx 232$~km/s is the speed of the Sun in the halo rest frame.

The XENONnT experiment \cite{XENONnT} reported null result in the searches for dark matter particles with the mass ranging from 1 to 30 keV. The reported background is ($15.8\pm 1.3$) events/(ton$\times$year$\times$keV). This means that the scalar field absorption rate (\ref{rate2}) should not exceed the error in the reported background,
$R<R_\text{error}=1.3$  events/(ton$\times$year$\times$keV). This allows us to find the following limit on the coupling constant:
\begin{equation}
    |g_{\phi n}| < \left(
    \frac{4\pi\alpha R_\text{error} m_\text{at} m_\phi m_p^2}{\sigma_\gamma \langle v^2 \rangle \rho_\text{DM}m_e^2}\right)^{1/2}\,.
    \label{22}
\end{equation}
Assuming that the local DM energy density is $\rho_\text{DM} = 0.3\text{\,GeV}/\text{cm}^3$, we find the following limit at $m_\phi=3$\,keV:
\begin{equation}
    |g_{\phi n}|< 1.2\times 10^{-10}\,.
\end{equation}
Here we used the following value for the photoionization cross section $\sigma_\gamma = 1.55\times 10^{-19}\,\text{cm}^2$ at $\omega = 3$\,keV. 

\begin{figure}
    \centering
    \includegraphics[width=8cm]{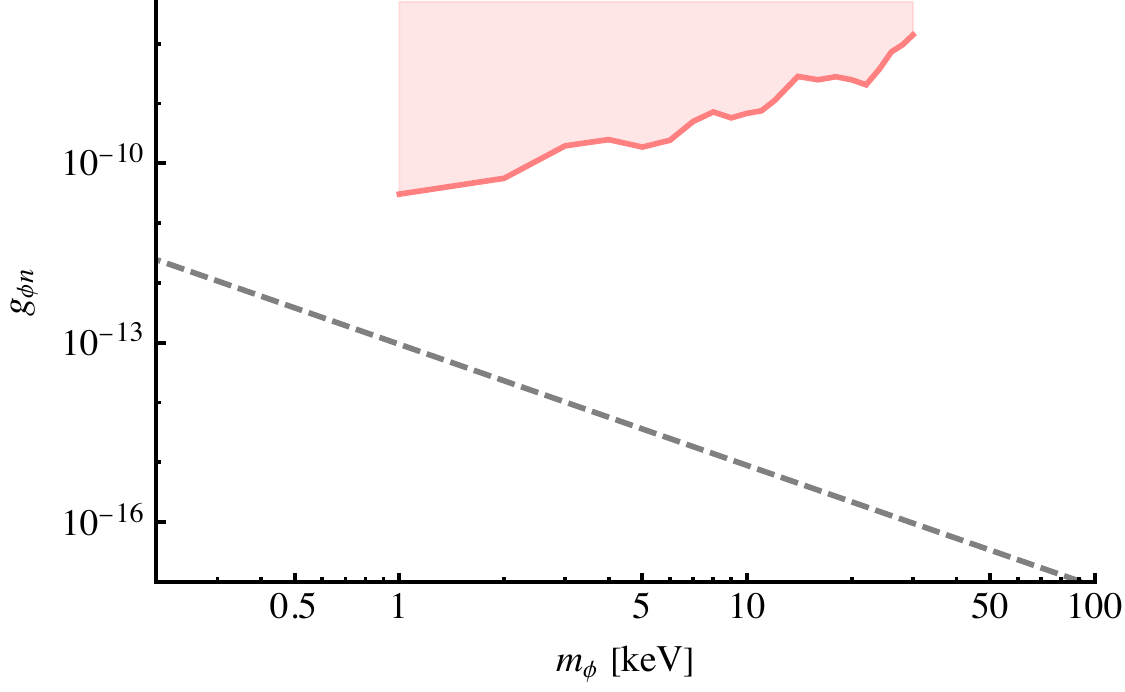}
    \caption{Pink excluded region for the scalar-nucleon coupling constant $g_{\phi n}$ is found from non-observation of DM particles in XENONnT experiment \cite{XENONnT}. Dashed line represents indirect limits on this constant from Ref.~\cite{myRecent}.}
    \label{fig:ExclusionPlot}
\end{figure}

To provide a more accurate estimate of the limits on the scalar-nucleon interaction, one has to consider the predicted transition rate (\ref{rate1}) on top of the actual background model in the XENONnT experiment \cite{XENONnT}, and compare it with the experimental data. The same result may be achieved by re-scaling the limits on the axion-electron coupling $g_{ae}$ from Ref.~\cite{XENONnT} using the relation
\begin{equation}
    \frac{\sigma_\phi}{\sigma_a} = \frac43 \frac{g_{\phi n}^2}{g_{ae}^2}
    \frac{v^2}{c^2}
    \frac{m_e^4}{m_p^2 m_\phi^2}\,,
\end{equation}
where $\sigma_\phi$ is given by Eq.~\eqref{sigma-phi} and $\sigma_{a}$ is the axio-ionization cross section calculated in Refs.~\cite{Pospelov2010,AxioElectric}. The corresponding limits on the scalar-electron coupling are given in Fig.~\ref{fig:ExclusionPlot} for mass ranging from 1 to 30 keV. Note that unlike the QCD axion model, there is no universal relation between the coupling constant $g_{\phi n}$ and the scalar particle mass $m_\phi$ which could indicate target sensitivity for experiments.

Recall that the XENONnT experimental data have a lower cutoff at 1 keV for the deposited energy. As a result, the measured cross section $\tilde\sigma_\phi$ in this experiment is significantly different from the theoretical cross section $\sigma_\phi$ near the 1 keV cutoff. The cross section $\tilde\sigma_\phi$ may be calculated numerically using the same formula (\ref{sigma-phi}), but one should keep in the sum only those electron shells which are characterized by the energy exceeding the experimental threshold in absolute value, $|{\cal E}_\text{i}|>1$\,keV. The results of these numerical calculations are given in the Appendix; the corresponding plots of these cross sections are shown in Fig.~\ref{fig:CrossSections} by a black dashed curve.

In Fig.~\ref{fig:ExclusionPlot}, we included indirect limits on the scalar-nucleon coupling from Ref.~\cite{myRecent} represented by dashed line. These limits are based on one-loop radiative corrections to the scalar-photon interaction. The latter coupling is constrained in the keV region by non-observation of excess in the diffuse x-ray background in various space-based x-ray observatories, see, Refs.~\cite{Xrays,XMM,NuStar}. Although  our limits are a few orders in magnitude weaker than the indirect ones, they should be considered as complimentary results because these two approaches are based on different effects.

\section{Summary}
In this paper, we propose new mechanism for atomic ionization by dark matter particles. We consider a model of dark matter represented by a real scalar field with $O$(keV) mass and Yukawa-type interaction to electron and nucleon. We assume that scalar-nucleon interaction is much stronger than the scalar-electron one, as the opposite case was studied in Ref.~\cite{Bao2021}. We derive a general formula (\ref{sigma-phi}) for absorption of this scalar field in atoms and show that it is strongly dominated by the dipole matrix element, in contrast with the case of scalar-electron interaction where the $E0$ atomic transitions give significant contributions to the cross section \cite{Bao2021}. Eq.~(\ref{sigma-phi}) is applicable for both cold dark matter particles and relativistic scalars which may be emitted by the Sun.

We have found a relation between the scalar particle absorption cross section in atoms to the photoionization cross section at the corresponding energy, see Eq.~(\ref{result}). Note that the number of excited electrons produced by absorption of one particle may be bigger than one. This relation allows one to consider the absorption of hypothetical scalar particles not only on isolated atoms, but also in liquid noble gas and solid state detectors. 

We also calculate the  dipole matrix element numerically for a variety of atoms of experimental interest: Na, Si, Ar, Ge, I, Xe, and Tl. This matrix element enters the expression of the scalar field absorption cross section on atoms (\ref{sigma-phi}). In the Appendix, we present tables of the results of numerical calculations of this cross section in the case of a massless scalar particles with energy ranging from 0.1 to 100 keV and analytical formulae fitting these numerical data. Files with detailed  numerical data are provided in the Supplemental Material. For massive particles there is extra factor $v/c$. These results may be used in the DM-searching experiments based on the liquid noble gas and solid state detectors. 

As a demonstration, we employ the null result in the XENONnT experiment \cite{XENONnT} in order to find a limit on the scalar-nucleon coupling constant, $|g_{\phi n}|<1.2\times 10^{-10}$ at $m_\phi=3$\,keV. For other energies, the exclusion plot is given in Fig.~\ref{fig:ExclusionPlot}. To the best of our knowledge, this is the first direct limit on this coupling constant for the scalar field DM with mass ranging from 1 to 30 keV from laboratory experiment. In the recent work \cite{myRecent}, done after submission of this paper, we complemented these laboratory limits by the ones obtained from astrophysical measurements of scalar particle decays into two photons.   

In conclusion, we would like to note that the suggested effect  is of interest from a theoretical physics perspective. We consider dark matter interaction with a nucleus. When considering such interaction, a bare nucleus, due to its  substantial mass $M$ relative to that of a dark matter particle,   receives a very small amount of kinetic energy, $E =q^2/2M$, from the dark matter particle; here $q$ is the momentum transfer. For example, $^{132}$Xe nucleus  receives kinetic energy of about $10^{-9}$\,eV from a non-relativistic dark matter particle with mass 10 keV. Contrastingly,  while dark matter interacts with atomic nucleus only, electron may receive energy exceeding 1 keV,  which is 12 orders of magnitude greater! Notably, all nucleons interact coherently with the DM particle, preserving the nucleus in its ground nuclear state. Thus, in this phenomenon, the entirety of the dark matter particle's energy, including its rest mass term, is directed towards the ionization of atomic electrons. Although the probability of such process is low, suppressed by the factor $(m_e/M)^2$, it is distinctly non-zero.   This phenomenon is fundamentally enabled by the quantum mechanical properties of atomic structure.

\vspace{2mm}
\textit{Acknowledgements.}--- 
We are grateful to Hoang Bao Tran Tan for useful discussions. The work was supported by the Australian Research Council Grants No.\ DP230101058 and No.\ DP200100150.


%

\newpage
\appendix
\section{Analytical formula for ionization cross section} 
\label{AppA}

The ionization cross section for photons and scalars  is dominated by the contribution from deepest shells allowed by the energy conservation. When the particle possesses sufficient energy for ionization from $2p_{3/2}$ electron orbital (see Table \ref{tab:2pEnergies}), the dipole term in the ionization cross section may be approximated by the analytical formula from  Ref.~\cite{Pospelov2010}. This formula includes contributions from $1s$, $2s$ and $2p$ atomic shells which were obtained by fitting the results of the numerical calculations,
\begin{equation}
    \sigma_\gamma({\cal E}) = \frac43 \pi\alpha\left(\frac{{\cal E}_0}{{\cal E}} \right)^2 K({\cal E})a_B^2\,,
    \label{SigmaGammaFit}
\end{equation}
where $a_B$ is the Bohr radius and the function $K$ is given by
\begin{align}
    K &= K_{1s} + K_{2s} + K_{2p}\,,\\
    K_{1s} &= f_1(Z,{\cal E}+{\cal E}_{1s})
    \frac{384 \pi {\cal E}_{1s}^4}{({\cal E}_0 Z {\cal E})^2}
    \frac{e^{-4\nu_1 \text{arccot} \nu_1}}{1-e^{-2\pi \nu_1}}\,,\label{K1s}\\
    K_{2s} &= f_2(Z,{\cal E}+{\cal E}_{2s})\frac{6144\pi e_2^3}{{\cal E}_0 {\cal E}^2}
    \left(1+3\frac{e_2}{\cal E} \right)\nonumber\\
    &\times\frac{e^{-4\nu_2\text{arccot}(\nu_2/2)}}{1-e^{-2\pi\nu_2}}\,,\label{K2s}\\
    K_{2p} &= f_2(Z,{\cal E}+{\cal E}_{2p}) \frac{12288\pi e_3^4}{{\cal E}_0 {\cal E}^3} 
    \left(3+8\frac{e_3}{\cal E} \right) \nonumber \\
    &\times \frac{e^{-4\nu_3 \text{arccot}(\nu_3/2)}}{1-e^{-2\pi\nu_3}}\,,
    \label{K2p}
\end{align}
where ${\cal E}_0 = 27.21$\,eV is the Hartree unit of energy, $e_2 = |{\cal E}_{2s}|$, $e_3 = |{\cal E}_{2p}|$, $\nu_1 = \sqrt{-{\cal E}_{1s}/({\cal E}_{1s}+{\cal E})}$, $\nu_2 = 2\sqrt{-{\cal E}_{2s}/({\cal E}_{2s}+{\cal E})}$, $\nu_3 = 2\sqrt{-{\cal E}_{2p}/({\cal E}_{2p}+{\cal E})}$. Here ${\cal E}_{1s}$, ${\cal E}_{2s}$ and ${\cal E}_{2p}$ are Hartree-Fock energies of the core states represented by the following fitting formulas for atoms with $18<Z<60$:
\begin{align}
    \frac{{\cal E}_{1s}}{{\cal E}_0} &= -\frac{Z^2 - 7.49 Z + 43.39}{2} \,,\\
    \frac{{\cal E}_{2s}}{{\cal E}_0} &= -0.000753 Z^3 - 0.028306 Z^2 \nonumber \\& -0.066954 Z + 2.359052\,,\\
    \frac{{\cal E}_{2p}}{{\cal E}_0} &= -0.000739 Z^3 - 0.027996 Z^2 \nonumber \\& +0.128526Z  +1.435129\,.
\end{align}
The functions $f_1$ and $f_2$ in Eqs.~(\ref{K1s},\ref{K2s},\ref{K2p}) are:
\begin{align}
    f_1(Z,{\cal E}) &=(5.368\times 10^{-7}Z - 1.17\times 10^{-4}){\cal E}/{\cal E}_0
    \nonumber\\& -0.012 Z + 1.598\,,\\
    f_2(Z,{\cal E}) &=(-1.33\times 10^{-6}Z + 1.17\times 10^{-4}){\cal E}/{\cal E}_0 \nonumber\\ &- 0.0156Z + 1.15\,.
\end{align}
Note that the approach in derivation of these formulas is based on correcting  the formulas obtained for ionization of electron in the pure Coulomb potential. 

\begin{table}[tb]
    \centering
    \begin{tabular}{c|c|c}
        $Z$ & atom & $|{\cal E}_{2p_{3/2}}|$, keV  \\\hline
        11 & Na & 0.04122 \\
        14 & Si & 0.1156 \\
        18 & Ar & 0.2598 \\
        32 & Ge & 1.528 \\
        53 & I  & 4.615 \\
        54 & Xe & 4.835 \\
        81 & Tl & 12.76
    \end{tabular}
    \caption{Hartree-Fock energies of the $2p_{3/2}$
electron shells of some neutral atoms of interest.}
    \label{tab:2pEnergies}
\end{table}

Making use of Eq.~(\ref{SigmaGammaFit}), the cross section (17) may be cast in the form 
\begin{equation}
    \sigma_\phi({\cal E}) = \frac{g_{\phi n}^2}{3}\frac{m_e^2}{m_p^2}\frac{k}{\cal E}\left(\frac{{\cal E}_0}{{\cal E}} \right)^2 K({\cal E})a_B^2\,.
    \label{A11}
\end{equation}
This formula is plotted in Fig.~1 by a green line labeled by ``Fit.''  It shows a good agreement with the results of direct numerical calculations for the cross section (dashed orange curve) and with the use of experimental values of photoionization cross section (solid blue curve).


\section{Numerical results for cross section of atomic ionization through scalar field absorption}
\label{AppB}

The general formula for cross section of atomic ionization with absorption of a scalar particle with scalar-nucleon interaction is given by Eq.~(14). This cross section is expressed via the standard electric dipole matrix element that may be calculated numerically for isolated atoms to a high accuracy. We calculated this matrix element using relativistic Hartree-Fock method for a variety of atoms of experimental interest, Na, Si, Ar, Ge, I, Xe, and Tl, for the scalar field with the energy ranging from 0.1 to 100 keV. In Table~\ref{tab:sigma-phi}, we provide the results of calculations of the absorption cross section obtained with a massless analog of Eq.~(14),
\begin{equation}
    \frac{\sigma_\phi(\omega)}{g_{\phi n}^2} = \omega \frac{m_e^2}{m_p^2} \frac\pi3 \sum_{\text{i,f}} |\langle \text{f} | \vec R |\text{i}\rangle |^2\,.
    \label{sigma-appendix}
\end{equation}
Here $|\text{i}\rangle$ is the ground state of the atom, and $|\text{f}\rangle$ is an energy eigenstate of the ionized electron in the Coulomb field of the ion.

In the case of non-relativistic galactic halo DM particles, the corresponding cross section is obtained from Eq.~(\ref{sigma-appendix}) by a re-scaling with the factor $v/c$ (for DM $v/c \sim 10^{-3}$). 

It is important to note that some of the DM particle detectors have lower energy threshold for detection of ionized electrons through photo-scintillation. For instance, the detectors in the XENONnT experiment \cite{XENONnT} have the electron energy cut-off at ${\cal E}_\text{cut} =1$\,keV. This means that the application of the formula (\ref{sigma-appendix}) for estimates of the event rate in such detectors would give an overestimated result at the energies near the threshold. 

\begin{figure*}
    \centering
    \includegraphics[width=8cm]{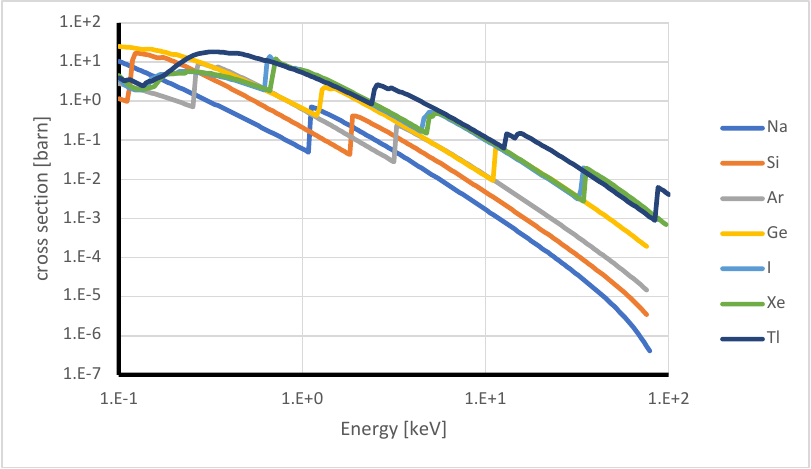}
    \includegraphics[width=8cm]{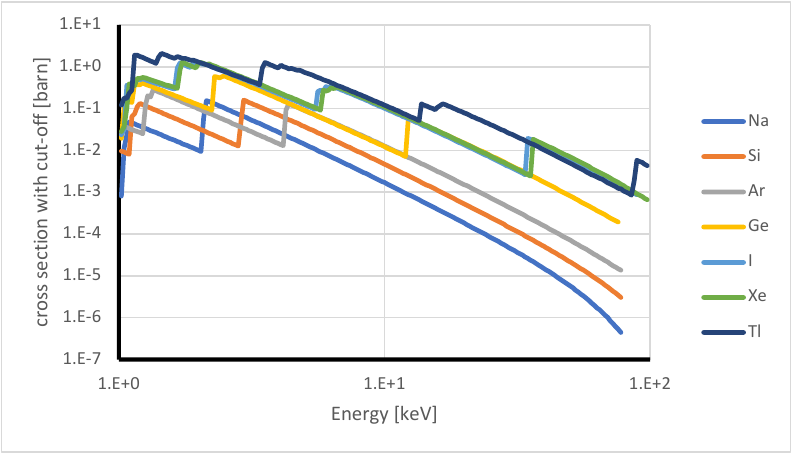}
    \caption{Left: cross sections $\sigma_\phi/g_{\phi n}^2$ of atomic ionization with absorption of a massless scalar field calculated numerically with Eq.~(14). Right: similar cross sections, but assuming the energy detection threshold for ionized electrons at 1 keV.}
    \label{fig:allCrossSections}
\end{figure*}

In order to take into account the energy threshold of ionized electrons in actual detectors, we consider a modified cross section, $\tilde\sigma_\phi$, which is formally defined by the same expression (\ref{sigma-appendix}), but the sum should include only those electron shells which are characterized by the energy exceeding the detector threshold in absolute value, $|{\cal E}_\text{i}|>{\cal E}_\text{cut}$. In Table \ref{tab:sigma-tilde-phi}, we tabulate the results of numerical calculations of $\tilde\sigma_\phi$ assuming the energy cut-off at ${\cal E}_\text{cut}=1$\,keV. The cross sections $\sigma_\phi$ and $\tilde\sigma_\phi$ are plotted in Fig.~\ref{fig:allCrossSections}.

\onecolumngrid

\begin{table*}[tb]
\scriptsize
    \centering
    \begin{tabular}{c|c|c|c|c|c|c|c}
 & \multicolumn{7}{c}{$\sigma_\phi/g_{\phi n}^2$, barn}\\\hline
$\omega$, keV & Na & Si & Ar & Ge & I & Xe & Tl \\\hline
 0.2000 & 2.591 & 10.44 & 1.068 & 15.82 & 5.101 & 5.166 & 7.292 \\
 0.2636 & 1.427 & 5.898 & 5.367 & 10.84 & 5.522 & 5.710 & 15.71 \\
 0.3474 & 0.7691 & 3.129 & 7.520 & 6.623 & 4.435 & 4.723 & 18.29 \\
 0.4578 & 0.4077 & 1.604 & 4.261 & 3.830 & 3.137 & 3.395 & 16.56 \\
 0.6034 & 0.2121 & 0.7988 & 2.245 & 2.128 & 2.033 & 2.216 & 11.56 \\
 0.7953 & 0.1083 & 0.3921 & 1.135 & 1.142 & 8.183 & 8.594 & 8.039 \\
 1.048 & 0.05418 & 0.1902 & 0.5585 & 0.5992 & 5.653 & 6.002 & 4.980 \\
 1.381 & 0.4707 & 0.09108 & 0.2682 & 2.188 & 3.225 & 3.465 & 2.881 \\
 1.821 & 0.2388 & 0.04299 & 0.1276 & 1.353 & 1.690 & 1.823 & 1.592 \\
 2.399 & 0.1151 & 0.2707 & 0.06016 & 0.6741 & 0.8535 & 0.9254 & 0.8533 \\
 3.162 & 0.05345 & 0.1330 & 0.02805 & 0.3245 & 0.4219 & 0.4590 & 2.039 \\
 4.168 & 0.02415 & 0.06242 & 0.1463 & 0.1531 & 0.2021 & 0.2201 & 1.168 \\
 5.493 & 0.01065 & 0.02838 & 0.07067 & 0.07004 & 0.5014 & 0.4754 & 0.5902
   \\
 7.239 & 0.004620 & 0.01257 & 0.03279 & 0.03172 & 0.2472 & 0.2644 & 0.2897
   \\
 9.541 & 0.001968 & 0.005450 & 0.01474 & 0.01419 & 0.1166 & 0.1254 & 0.1388
   \\
 12.57 & 0.0008244 & 0.002317 & 0.006445 & 0.05453 & 0.05387 & 0.05805 &
   0.06373 \\
 16.57 & 0.0003385 & 0.0009681 & 0.002752 & 0.02568 & 0.02388 & 0.02595 &
   0.1317 \\
 21.84 & 0.0001352 & 0.0003966 & 0.001148 & 0.01151 & 0.01038 & 0.01125 &
   0.06175 \\
 28.79 & 0.00005172 & 0.0001581 & 0.0004694 & 0.004952 & 0.004408 &
   0.004782 & 0.02787 \\
 37.94 & 0.00001841 & 0.00006067 & 0.0001866 & 0.002051 & 0.01508 & 0.01616
   & 0.01214 \\
 50.00 & $5.808\times 10^{-6}$ & 0.00002172 & 0.00007158 & 0.0008176 & 0.006448
   & 0.006957 & 0.004975
    \end{tabular}
    \caption{Results of numerical calculation of cross section (14) using relativistic Hartree-Fock method in the case of massless scalar field. More detailed data tables are provided in separate files.}
    \label{tab:sigma-phi}
\end{table*}

\begin{table*}[tb]
\scriptsize
    \centering
    \begin{tabular}{c|c|c|c|c|c|c|c}
 & \multicolumn{7}{c}{$\tilde\sigma_\phi/g_{\phi n}^2$, barn}\\\hline
$\omega$, keV & Na & Si & Ar & Ge & I & Xe & Tl \\\hline
 1.000 & 0. & 0. & 0. & 0. & 0. & 0. & 0. \\
 1.216 & 0.03700 & 0.1281 & 0.02587 & 0.4195 & 0.5292 & 0.5236 & 1.712 \\
 1.479 & 0.02222 & 0.07576 & 0.2235 & 0.2600 & 0.3547 & 0.3850 & 1.979 \\
 1.798 & 0.01337 & 0.04447 & 0.1320 & 0.1596 & 1.062 & 1.174 & 1.558 \\
 2.187 & 0.1476 & 0.02594 & 0.07753 & 0.09828 & 1.077 & 1.167 & 1.055 \\
 2.659 & 0.08678 & 0.01502 & 0.04535 & 0.5151 & 0.6572 & 0.7140 & 0.6682 \\
 3.234 & 0.05017 & 0.1253 & 0.02635 & 0.3057 & 0.3982 & 0.4330 & 0.4203 \\
 3.932 & 0.02862 & 0.07342 & 0.01520 & 0.1798 & 0.2360 & 0.2575 & 0.9690 \\
 4.782 & 0.01612 & 0.04229 & 0.1024 & 0.1037 & 0.1398 & 0.1524 & 0.8310 \\
 5.815 & 0.008982 & 0.02403 & 0.06049 & 0.05953 & 0.2662 & 0.08983 & 0.5107
   \\
 7.071 & 0.004965 & 0.01349 & 0.03506 & 0.03395 & 0.2626 & 0.2814 & 0.3081
   \\
 8.599 & 0.002718 & 0.007482 & 0.01999 & 0.01924 & 0.1551 & 0.1667 & 0.1832
   \\
 10.46 & 0.001475 & 0.004111 & 0.01123 & 0.01084 & 0.09039 & 0.09728 &
   0.1080 \\
 12.72 & 0.0007961 & 0.002237 & 0.006229 & 0.05298 & 0.05217 & 0.05624 &
   0.06175 \\
 15.46 & 0.0004240 & 0.001206 & 0.003414 & 0.03121 & 0.02943 & 0.03199 &
   0.09419 \\
 18.8 & 0.0002232 & 0.0006454 & 0.001850 & 0.01790 & 0.01636 & 0.01772 &
   0.09356 \\
 22.87 & 0.0001157 & 0.0003412 & 0.0009916 & 0.01003 & 0.009021 & 0.009777
   & 0.05424 \\
 27.81 & 0.00005859 & 0.0001780 & 0.0005258 & 0.005517 & 0.004914 & 0.005330
   & 0.03080 \\
 33.81 & 0.00002871 & 0.00009105 & 0.0002750 & 0.002974 & 0.002646 & 0.00287
   & 0.01724 \\
 41.12 & 0.00001336 & 0.00004542 & 0.0001420 & 0.001574 & 0.01188 & 0.01276
   & 0.009444 \\
 50.00 & $5.793\times 10^{-6}$ & 0.00002173 & 0.00007157 & 0.0008177 & 0.006448
   & 0.006958 & 0.004975 
    \end{tabular}
    \caption{Results of numerical calculation of cross section (14) in the case of massless scalar field assuming a threshold of electron energy at 1 keV in the detector. More detailed data tables are provided in separate files.}
    \label{tab:sigma-tilde-phi}
\end{table*}

\end{document}